\documentclass{elsart}
\usepackage{graphicx,amssymb}

\begin{document}
\thispagestyle{empty}

\def\etal{{\sl et al.}}
\newcommand{\kdc}{$K^{\pm}\rightarrow \pi^{\pm} \pi^{0} \pi^{0}$}
\newcommand{\kpdc}{$K^{+} \rightarrow \pi^{+}\pi^{0}\pi^{0} $}
\newcommand{\kmdc}{$K^{-} \rightarrow \pi^{-}\pi^{0}\pi^{0} $}


\begin{frontmatter}
\author{G.A.~Akopdzhanov},
\author{V.B.~Anikeev},
\author{V.A.~Bezzubov},
\author{S.P.~Denisov},
\author{A.A.~Durum},
\author{Yu.V.~Gilitsky},
\author{V.M.~Korablev},
\author{V.I.~Koreshev},
\author{A.V.~Kozelov\thanksref{cora}},
\author{E.A.~Kozlovsky},
\author{V.I.~Kurbakov},
\author{V.V.~Lipaev},
\author{A.M.~Rybin},
\author{A.A.~Shchukin},
\author{M.M.~Soldatov},
\author{D.A.~Stoyanova},
\author{K.I.~Trushin},
\author{I.A.~Vasilyev}

\address{State Research Center 
Institute for High Energy Physics, Protvino, 
Moscow region, 142281 Russia}
\thanks[cora]{Corresponding author. {\it E-mail address:} Alexander.Kozelov@ihep.ru}

\title{Measurements of the Dalitz Plot Parameters 
for $K^{\pm}\to\pi^{\pm}\pi^0\pi^0$ Decays }

\journal{Eur.\ Phys.\ J.\ C}

\begin{abstract}
The $g$, $h$, and $k$  Dalitz plot parameters, which are coefficients in a series expansion of the squared module of the matrix element 
$|M(u,v)|^{2} \propto 1 + gu + hu^{2} + kv^{2}$ 
($u$, $v$ are invariant variables), have been measured for \kdc\ decays using 
$35~GeV/c$ hadron beams at the IHEP (Protvino) accelerator. Dependences of parameters and fit quality on the $\pi^0\pi^0$ mass cut were investigated.
The results point to the important role of  $\pi^+\pi^- \rightarrow \pi^0\pi^0$
 charge exchange scattering near the $\pi^0\pi^0$ mass threshold. The comparison of our data with previous measurements is presented.
\end{abstract}


\end{frontmatter}

\section{Introduction}

      

The Dalitz plot parameters for \kdc\ decays measured in the three most precise experiments
 with $K^-$ \cite{bolotov,ajinenko} and $K^+$ \cite{batusov} beams differ by 2 to 5
standard deviations
\cite{pdg}.  For example, the difference of the Dalitz plot slopes $g$ 
obtained in \cite{ajinenko} and \cite{batusov} is equal to $0.109\pm0.021$.
 As shown by our data \cite{dghk} this result cannot be explained by CP violation and is most probably due to underestimation of systematic uncertainties. In this paper we present new results on the Dalitz plot parameters based on the analysis of 493k events of \kdc\ decays collected with the TNF-IHEP facility \cite{dghk,proposal}.

%

\section{Experimental Setup}

      Studies of charged kaon decays have been performed using $35~GeV/c$ positive and negative hadron beams at the IHEP accelerator. The beam intensity was monitored by four scintillation counters. Its typical value was $4 \cdot 10^6$ per $1.7$ second spill. Three threshold and 
two differential Cherenkov counters were used to select kaons with a background of less than 1\%. The products of kaon
decays originating in the 58.5 $m$ long vacuum pipe were detected by wide aperture scintillation hodoscopes and the total absorption electromagnetic
calorimeter GEPARD consisting of 1968 lead-scintillator cells.  The $\pi^0$ mass resolution was  
$12.3~MeV/c^2$. The calorimeter
was divided into 16 trigger elements. An anticoincidence beam counter was  placed downstream of the vacuum pipe.
       The first level trigger T1 was formed according to the following logic formula:
 $$T1=S1\cdot S2\cdot S3\cdot S4\cdot (D1+D2)\cdot\overline{C1}
\cdot\overline{C2}\cdot\overline{C3}\cdot\overline{AC},$$ 
where S$i$, $Di$, $Ci$, and $AC$ are logical signals from the beam, differential, threshold, and anticoincidence counters respectively. The Level 2 trigger required more than 0.8~$GeV$ energy deposition in at least three trigger elements of the GEPARD.
      The details of the setup and measurement procedure 
can be found elsewhere \cite{dghk}.

\section{\kdc\ event selection}

The following criteria were used to select \kdc\ events \cite{dghk}:
\begin{itemize}
\item one to three secondary tracks are reconstructed;
\item the probability of the decay vertex fit is more than 5\%;
\item the decay vertex is inside the fiducial length of the decay pipe;
\item the number of clusters with energy above 1~$GeV$ in the calorimeter and 
the number of tracks in the hodoscopes correspond to the \kdc\ decay;
\item charged pion energy exceeds 8~$GeV$;
\item the $\chi^2$ probability $P(\chi^2)$ of the 6C kinematic fit is more than 0.1 
(all possible photon combinations are considered and the best is selected);
\item event passes software Level~2 trigger.
\end{itemize}

The experimental setup was simulated using the Monte Carlo (MC) method with the GEANT~3.21 code. 
The setup geometry was described in detail and the data 
obtained in the experiment were taken into account.
Among these data there were calibration coefficients for each channel of the 
calorimeter, the dependence of the hodoscope efficiency on the particle coordinates and correlations between kaon's spatial and angular coordinates 
and its momentum. 

The final data sample includes 493K completely reconstructed \kdc\  events.
The MC statistics is about four times higher.
The background level estimated
from the MC simulations is less than 0.25\% and is mainly due to 
$K^{\pm}\rightarrow \pi^{\pm}\pi^0$ decays.
It is shown in our paper \cite{dghk} that the event distributions in the Dalitz plots
 for  $K^+$ and $K^-$ decays are identical.  
Taking this into account we used combined statistics for \kdc\ decays 
 to estimate the Dalitz plot parameters. 

\section{Results}

\begin{table*}
\caption{Fit results with higher-order terms in (\ref{eq.ghkdef})}
\label{table.higher}
\begin{center}
\begin{tabular}{|c|c|c|c|c|}

  \hline
 $g$ & $0.6259\pm0.0043$ & $0.6151\pm0.0051$ & $0.6284\pm0.0048$ & $0.6129\pm0.0063$\\
  \hline
 $h$ & $0.0551\pm0.0044$ & $0.0782\pm0.0073$ & $0.0556\pm0.0044$ & $0.0795\pm0.0077$\\
  \hline
 $k$ & $0.0082\pm0.0011$ & $0.0080\pm0.0011$ & $0.0070\pm0.0015$ & $0.0087\pm0.0016$\\
   \hline
 $l$ &     ---           & $0.0273\pm0.0069$ &      ---          & $0.0292\pm0.0076$\\
   \hline
 $m$ &    ---            &   ---             & $-0.0027\pm0.0024$& $0.0016\pm0.0027$ \\
   \hline
 $\chi^2$ & 506.1 &490.2 & 504.8 & 489.8 \\
   \hline
  $\chi^2/ndf$& 1.18 & 1.14 & 1.17 &  1.14 \\
  \hline
  $P(\chi^2)$ &$6.6\cdot10^{-3}$& $2.4\cdot10^{-2}$ &$7.4\cdot10^{-3}$&$2.4\cdot10^{-2}$\\
   \hline
\end{tabular}
\end{center}
\end{table*}

The following  parametrization  of the squared module of the matrix element
for \kdc\ decays was
used in the data analysis \cite{pdg}:
\begin{eqnarray} 
|M(u,v)|^{2} \propto 1 + gu + hu^{2} + kv^{2} 
\label{eq.ghkdef}
\end{eqnarray}

Due to the finite setup resolution on the $u$ and $v$ variables \cite{dghk} the
 `measured' $u'$, $v'$ values can differ from the true $u$, $v$ for both 
experimental and MC events.  To take this into account the Dalitz plot 
parameters were estimated by minimizing the following functional form: 

\[
\chi ^2(g,h,k)  = \sum\limits_{i}^{Nbin} {\frac{{\left( {n_{i}  - C \cdot m_{i})}  \right)^2 }}{{\sigma_{i} ^2 }}}, 
\]
where $n_{i}$ is the number of events in the $i$-th experimental Dalitz plot bin,
$ m_i  \equiv m_i (g,h,k) = \sum\limits_j  {w_{ij} }$
$ (w_{ij}  = 1 + g \cdot u_j  + h \cdot u_j^2  + k \cdot v_j^2 )$
is a sum of the weighted MC events in the $i$-th Dalitz plot bin,
$C = {{\sum {n_i } } \mathord{\left/
 {\vphantom {{\sum {n_i } } {\sum {m_i } }}} \right.
 \kern-\nulldelimiterspace} {\sum {m_i } }}$
is a normalization factor and
$ \sigma _i^2  = n_i  + C^2 \cdot \sum\limits_j {w_{ij}^2 } $
takes into account 
 limited MC statistics. The following values of the $g$, $h$, $k$ parameters and  elements of the correlation matrix were obtained: 

\begin{equation}
\left\{ \begin{array}{lll}
g & = & 0.6259 \pm 0.0043,  \\
h & = & 0.0551 \pm 0.0044, \\
k & = & 0.0082 \pm 0.0011,
\end{array} \right.
\qquad
\left( \begin{array}{ccc}
 1.00 &  0.90 &  0.41 \\
      &  1.00 &  0.33 \\
      &       &  1.00
\end{array} \right).
\label{eq.result}
\end{equation}

\begin{figure*}
\begin{center}
\includegraphics*[scale=0.7,bb=50 20 300 300]{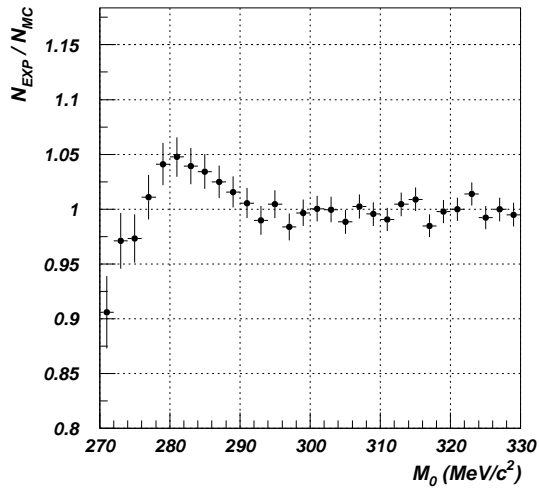} 
\end{center}
\caption{Ratio of experimental to MC events vs
$\pi^0\pi^0$ invariant mass $M_{0}$. 
}
\label{fig:ratio}
\end{figure*}

The errors quoted are statistical only, 
 $\chi^2/{\it ndf}$ is 506/430 = 1.18 and $P(\chi^2)=0.0066$.
The low significance of the fit is primarily due to
a difference between the experimental data 
and the MC simulations based on equation 
(\ref{eq.ghkdef}) in the threshold region of the $\pi^0\pi^0$ invariant mass $M_0$
(Fig.~\ref{fig:ratio}).
This discrepancy can not be avoided by the addition of higher order terms 
$l\cdot u^3$ and/or $m\cdot uv^2$ in the $|M(u,v)|^{2}$ expansion 
(see Table~\ref{table.higher})
and might be due to nonanalitical terms in the matrix element 
connected to $\pi^+\pi^-\rightarrow\pi^0\pi^0$ rescattering \cite{likhoded}.
This effect was recently considered  in detail by Cabibbo \cite{cabibbo1} and
Cabibbo, Isidory \cite{cabibbo2}.
 It plays an important role  in the region of
$ M_0   \sim 2m_{\pi ^0 } $
and its contribution can be suppressed by introducing a complementary criterion of $M_0>M_{T}$.
It appeared that the fit with $M_T \ge 290$~$MeV/c^2$ results in stable values
of the Dalitz plot parameters independent from the $M_0$ cut and a satisfactory fit
significance. Introduction of the higher order terms in Eq.(\ref{eq.ghkdef})
does not change the $g$, $h$, $k$ parameters and $\chi^2$ value in this case.
The fit with $M_T=290$~$MeV/c^2$ gave the following results: 

\begin{equation}
\left\{ \begin{array}{lll}
g & = & 0.6339 \pm 0.0046,  \\
h & = & 0.0593 \pm 0.0088, \\
k & = & 0.0083 \pm 0.0013,
\end{array} \right.
\qquad
\left( \begin{array}{ccc}
 1.00 &  0.52 &  0.43 \\
      &  1.00 &  0.16 \\
      &       &  1.00
\end{array} \right),
\label{eq.result2}
\end{equation}
and $\chi^2/ndf=1.04$, $P(\chi^2)=0.3$.

\begin{figure*}
\begin{center}
\includegraphics*[scale=0.8,bb=50 20 450 500]{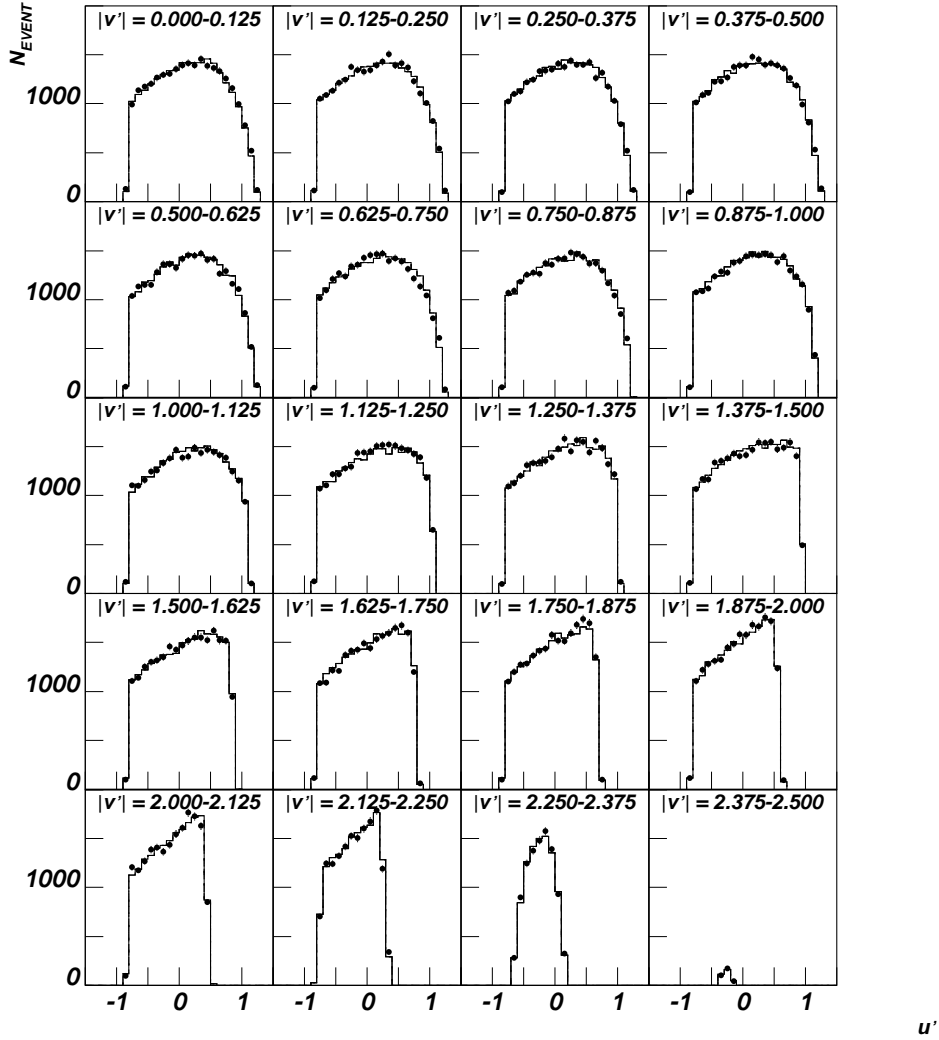} 
\end{center}
\caption{
 The $u'$ distribution of events in 
different intervals of $|v'|$
(histogram -- simulation, circles -- experiment).
}
\label{fig:uv2d}
\end{figure*}

\begin{figure*}
\begin{center}
\includegraphics*[scale=0.9,bb=20 20 510 180]{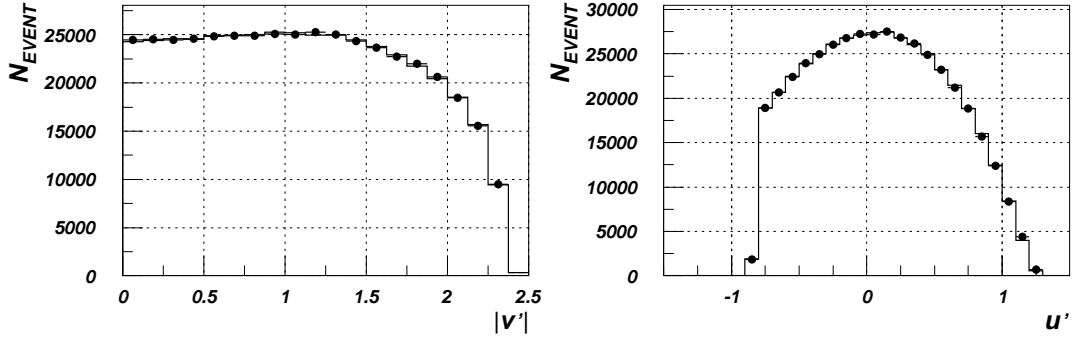} 
\end{center}
\caption{
Event distributions projected on the $u'$ and $|v'|$ axes 
(histogram - simulation, circles - experiment).
}
\label{fig:uvproj}
\end{figure*}

\begin{figure*}
\begin{center}
\includegraphics*[scale=0.7,bb=50 20 400 240]{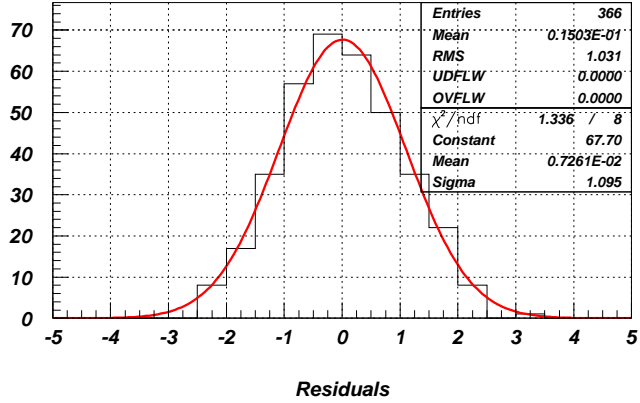} 
\end{center}
\caption{
Normalized residuals of the fit $(n_{i}-m_{i})/\sigma_{i}$.
}
\label{fig:resid}
\end{figure*}

Comparison of  the matrices in (\ref{eq.result}) and (\ref{eq.result2})
 shows that the correlations between $g$, $h$, and $h$, $k$ are much weaker
 (significantly lower) if the $M_T$ cut is applied. Figs.~\ref{fig:uv2d}-\ref{fig:resid} confirm the good agreement between experimental and MC data with
 $M_T=290$~$MeV/c^2$.


%

   To estimate the systematic uncertainties of the Dalitz plot parameters we checked the stability of the results against variation of the cuts in the event selection criteria. The parameters appeared to be most sensitive to the change of the minimum gamma energies  from 1 to 2~$GeV$ ($\Delta g=-0.0057$, $\Delta h=-0.0047$, $\Delta k=-0.0006$) and of the minimum charged pion momentum from 1 to 8~$GeV/c$ ($\Delta g=0.0048$, $\Delta h=0.0051$ and $\Delta k=0.0011$). The change in the bin size by factors of 2 and 0.5 and exclusion of the bins at the Dalitz plot boundary from the fit gives  $\Delta g=0.0012$, $\Delta h=0.0045$, and $\Delta k=0.0004$. Uncertainties in the kaon momentum, beam profile and angular spread, as well as GEPARD calibration coefficients have no influence on the parameters. The background contribution to systematic errors turned out to be negligible. Finally, our estimations of the systematic uncertainties are the following: $\delta g=0.0093$, $\delta h=0.0086$, $\delta k=0.0014$. They do not include the errors connected with $g$, $h$, $k$ variations due to introducing the $M_0$ cut 
 and the higher order terms in the expansion 
(\ref{eq.ghkdef}). 

\begin{figure*}
\begin{center}
\includegraphics*[scale=0.9,bb=20 30 500 200]{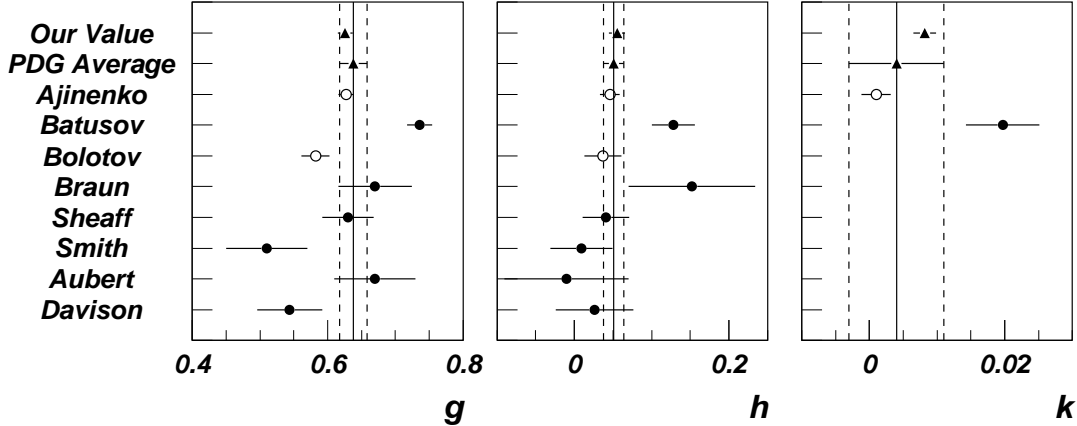} 
\end{center}
\caption{
The Dalitz plot parameters $g$, $h$ and $k$ for the 
\kpdc\ (solid circles),  \kmdc\ (open circles) and  \kdc\  (triangles) decays.
Vertical solid and dotted lines show the average values and their errors
as calculated by the PDG \cite{pdg}.
}
\label{fig:ghkcmp}
\end{figure*}

      Fig.~\ref{fig:ghkcmp} shows our results (\ref{eq.result}) together with previous measurements of the Dalitz plot parameters 
\cite{bolotov,ajinenko,batusov,pdg,davison,aubert,smith,sheaff,braun}.  
The error bars include both systematic and statistical uncertainties.

\section{Conclusions}

  The new data on the Dalitz plot parameters for \kdc\ decays
 based on the analysis of  $\sim 0.5M$ events collected with the TNF-IHEP facility are presented.
 The following results were obtained without a cut on the $\pi^0\pi^0$ invariant mass:
 $g=0.6259\pm 0.0043 \,(stat) \pm 0.0093 \,(syst)$,
 $h=0.0551\pm 0.0044 \,(stat) \pm 0.0086 \,(syst)$, 
 $k=0.0082\pm 0.0011 \,(stat) \pm 0.0014 \,(syst)$.
 The $g$ and $h$ values are in good agreement
 with those of Ajinenko {\it et al} \cite{ajinenko}.
 We observe a deviation of the $k$ value from zero $\sim 4.5$ standard deviations 
while Ajinenko {\it et al} reported $k=0.001 \pm 0.002$.
 We investigated the dependence of the Dalitz plot parameters and the fit quality on the $\pi^0\pi^0$ invariant mass cut $M_T$. It turned out that the fit significance
 becomes rather high if a cut of $M_T \ge 290$~$MeV/c^2$ is applied. 
 With this cut the addition of   higher order terms in expansion (\ref{eq.ghkdef})
 does not influence the  $g$, $h$, and $k$ parameters and the value of $\chi^2/ndf$.
 These results may be considered as evidence  of the important contribution
 of $\pi^+\pi^- \rightarrow \pi^0\pi^0$ rescattering \cite{cabibbo1,cabibbo2}
  to the matrix element of  the  \kdc\ decay in the threshold region 
  of the $\pi^0\pi^0$ invariant mass.

\section*{Acknowledgments}

     We appreciate the support of the experiment from A.A.Logunov, N.E.Tyurin, and A.M.Zaitsev and valuable contributions from V.N.Mikhailin, Yu.V.Mikhailov, V.A.Sen'ko, and A.N.Sytin. These studies were supported in part by the President grant 1305.2003.2 and by the Russian Fund for Basic Research (grants 05-02-16557 and 05-02-17614).

\newpage

\end{document}